\documentclass[10pt,conference]{IEEEtran}
\IEEEoverridecommandlockouts
\usepackage{cite}
\usepackage{amsmath,amssymb,amsfonts}
\usepackage{algorithmic}
\usepackage{graphicx}
\usepackage{textcomp}
\usepackage{xcolor}

\usepackage{dblfloatfix} 
\usepackage{subcaption} 

\def\BibTeX{{\rm B\kern-.05em{\sc i\kern-.025em b}\kern-.08em
    T\kern-.1667em\lower.7ex\hbox{E}\kern-.125emX}}
\begin{document}

\title{Open-Source Projects and their \\Collaborative Development Workflows}

\author{\IEEEauthorblockN{Panuchart Bunyakiati}
\IEEEauthorblockA{\textit{Department of Computer Engineering} \\
\textit{Kasetsart University}\\
Krungthep, Thailand \\
panuchart.b@ku.th}
\and
\IEEEauthorblockN{Usa Sammapun}
\IEEEauthorblockA{\textit{Department of Computer Science} \\
\textit{Kasetsart University}\\
Krungthep, Thailand \\
usa.sa@ku.th}
}

\maketitle

\begin{abstract}
For teams using distributed version control systems, the right collaborative development workflows can help maintaining the long-term quality of project repositories and improving work efficiency. Despite the fact that the workflows are important, empirical evidence on how they are used and what impact they make on the project repositories is scarce. Most suggestions on the use of workflows are merely expert opinions. Often, there is only some anecdotal evidence that underpins those advices. As a result, we do not know what workflows are used, how the developers are using them, and what impact do they have on the project repositories. This work investigates the various collaborative development workflows in eight major open-source projects, identifies and analyses the workflows together with the structures of their project repositories, discusses their similarities and differences, and lays out the future works. 
\end{abstract}


\section{Introduction}
In recent years, Git, a distributed version control system, has become one of the de facto standard tools for software development. Teams use Git as a medium for collaboration as well as a source code repository (repo) from where developers fork, clone, branch, commit, merge, push and pull changes. The repo has become a single source of truth for the software that all team members can refer to, rely on, access to and deploy from. And because of this, project repo must be maintained carefully, yet efficiently. Collaborative development workflows therefore play an important role in repository maintenance. 

To use Git effectively in a larger project, a team must establish a workflow which defines how the team members are to work together. Workflows, being the sequence of steps through which tasks are performed and work products are created, modified and used, are essential as they prescribe the convention which the team members follow and expect others to follow as well. Flout with this convention, the team may defile the repositories, create inconsistencies, find it difficult to track works and solve problems that arise later. When the workflow is right for the team, the long-term quality of project repositories increases. And works are performed more effectively. The wrong workflow hinders developers in their development efforts, may cause the developers to make unnecessary mistakes, and would eventually be circumvented or ignored.

\begin{table*}[ht]
	\caption{the general characteristics of open-source project repositories: as of April 2019}
	\label{projects}
	\centering
	\begin{tabular}{|l|r|r|r|r|r|r|r|} 
		\hline
Project 		& age (days) 	& contributors 	& commits 	& issues	& releases 	& merges (non-ff)	& reverts \\
\hline
Angular 		& 3,403		& 1,596 		& 8,962 		& 17,698	& 204 		& 27 (0.30\%)		& 124 (1.4\%)\\
\hline
jQuery 		& 3,680		& 273 		& 6,398 		& 1,844	& 149  		& 228 (3.56\%)		& 148 (2.3\%)\\
\hline
React 		& 2,163		& 1,291 		& 10,902 		& 7,369 	& 116 		& 3,051 (27.99\%)	& 82 (0.8\%)\\
\hline
Vue 			& 2,102		& 270 		& 3,010		& 7,948 	& 248 		& 141 (4.68\%)		& 22 (0.7\%)\\
\hline
.NET CoreFX 	& 1,637		& 773 		& 33,151		& 15,776 	& 81 			& 9,984 (30.12\%)	& 135 (0.4\%)\\
\hline
Node 		& 1,617		& 2,425 		& 26,901 		& 9,818 	& 544 		& 390 (1.45\%)		& 339 (1.3\%)\\
\hline
Rails 		& 4,037		& 3,805 		& 73,260  		& 11,593 	& 365 		& 19,848 (27.09\%)	& 1,189	(1.6\%)\\
\hline
Spring 		& 3,066		& 357 		& 18,497 		& 17,963 	& 150 		& 799 (4.32\%)		& 47 (0.3\%)\\
\hline
\end{tabular}
\end{table*}

\section{Workflows on Git}

Workflows on Git revolve around branching. A developer may create a branch to relate commits together for purposes such as developing a new feature, preparing a new release or fixing a bug. The branch can then be merged into other branches by ways of (1) rebasing and fast-forward merging, to pull new commits from the upstream branch and append the new commits after them to linearise the commit history, or (2) merging non-fast forward, which is simpler to perform, to create a new merge commit that merge two branches together and keep the structure and history of the branches intact. It is the branch usage, i.e. from which branch it is separated, how it is managed, into which branch it is merged, and how it is merged, at the local and remote repos, that defines much of the workflows. 

Git gives flexibility to software development by not enforcing a single, standardised workflow but allowing teams to establish their own workflows as they see fit. As a result, over the years, there is an increasing number of recommended workflows for using Git, for instance, Gitflow, trunk-based flow, GitHub flow, GitLab flow, etc. Despite their popularity, these generic workflows are merely baseline good practices rather than rigid procedures that developers must strictly follow. Teams may modify, add or remove steps in the workflows to create alternative workflows that suit their circumstances and the conditions in which the teams operate. It can be observed that many software projects take advantage of the flexibility Git provides and use their own workflows to maintain code. 

These variations of workflows are different in details including (1) how redundant efforts are prevented, (2) how quality of code is controlled, (3) how those efforts are integrated and organised, and (4) how different software releases are kept consistent. These aspects of the workflows are studied. For instance, Yu et al. \cite{Yu18} investigate the impact of distributed development efforts that may introduce duplicated pull-requests. Zhou et al. \cite{Zhou18} study fork-based development and create a tool to support management of those forks. In the area of quality control for code, Paixao et al. \cite{Paixao17} use data from four open-source systems to investigate whether contributors are aware of architectural changes. In the area of work integration, Rausch et al. \cite{Rausch17} examine development practices that impact the outcome of builds in continuous integration environments. And in releases consistency, the Tartarian tool \cite{Bunyakiati17} is developed to support cherry picking of commits in long-running, multi released software.

\section{Projects and Their Workflows}

We examine eight popular open-source projects including AngularJs, jQuery, React, VueJs, .NET Core libraries (CoreFX), Node, Ruby on Rails and Spring framework. The oldest project amongst these, Rails, started since April 2008. Node, the newest project, started in November 2014. As of April 2019, the number of contributors of these projects ranging from 270 (Vue) to 3,805 (Rails). The number of commits ranging from 3,010 (Vue) to 73,260 (Rails). The number of releases between 81 (.NET CoreFX) and 544 (Node). Table \ref{projects} shows the project's age (in days), the number of contributors, the number of commits, issues, and releases, together with the percentage of non-fast forward merge commits to all commits, and the percentage of revert commits to all commits.

Our initial survey from the contributors' guides of these projects shows that none of the project uses the generic workflows out of the box. Upon analysing the actual repos of these projects, we found that \textbf{Angular} and \textbf{Node} uses trunk-based workflow where feature branches are created for developing new features; when these features complete, the feature branches are rebased from the master branch to arrange the commits in linearised order and then merged back into the master branch. To release a version, a release branch is created as a major release branch; minor releases are created by tagging the commits in this release branch. \textbf{Vue} also uses a workflow that is similar Angular but uses develop branch to integrate changes from contributors. 

\textbf{jQuery} also uses trunk-based workflows but minor releases have their own branches and the commits in these minor release branches are also tagged. \textbf{React} seems to be using a mix between the \emph{Angular} and \emph{jQuery} workflows. A minor release is tagged on the major release branch when there is only one commit for that minor release. When a minor release has more than one commit, a release branch for that release is created. For \textbf{Spring framework}, a major release branch is separated from the master branch for each major release, a minor release branch is then separated from this major release branch and the commits are tagged. This is rather similar to those in \emph{jQuery} but different in that a pull request is rebased and then merged into the most recent release branch or the master branch where appropriate. When the pull request is merged into the most recent branch, the most recent branch is then non-fast forward merged into master branch to keep the two branches consistent. \textbf{Rails} and \textbf{Dot Net Core} use non-fast forward merge, instead of rebase, to integrate changes into the master branch and release branches depending on where the feature and bug-fixing branches are separated from, keeping the branch structure and commit history intact, as shown in Table \ref{projects} where their percentages of non-fast forward merge commits are significantly higher than those of other projects.

To generalise this, most of the projects use some variations of workflows similar to the trunk-based workflow which lets developers make changes and frequently push or make pull requests to the trunk and create release branches from the trunk to prepare for the releases. Each major release has its own release branch separated from the master branch. Each minor release is either tagged on the major release branch or has its own release branch. Commits that are made to the trunk can also be reapplied (cherry-picked) to these release branches, or vice versa. Furthermore, the workflows described in these projects also focus on testing, continuous integration, reviewing and static analysis, and specify the format of commit messages and the naming convention of branches. In addition, the workflows also suggest squashing a number of commits together into a logical unit of change in one commit to create a meaningful commit history. 

\section{Conclusion and Future Works}
In this work, we analyse the workflows used in eight major open-source projects and examine the commit history and structure of project repos. We observe that these workflows share the focus on four aspects of collaborative software development. To the best of our knowledge, our work is the first to discuss comparative collaborative development workflows. The closest to our work is \cite{Kalliamvakou15}, which examines the adoption of open source-style collaborative workflows in commercial software projects but only focuses on generic workflows. In our future work, we will seek to understand these four aspects deeper, in the context of comparative collaborative workflows. We maintain that software development teams should pay attention to the collaborative workflows and that the tasks related to these four aspects of the workflows should be well supported by tools. Our future works also aim to confirm our cross-project findings in this work with more empirical evidence and quantitative data to establish the best practices in collaborative development and standard workflows that support open collaboration.

\begin{figure*}[!b]
\centering
\includegraphics[scale=0.25]{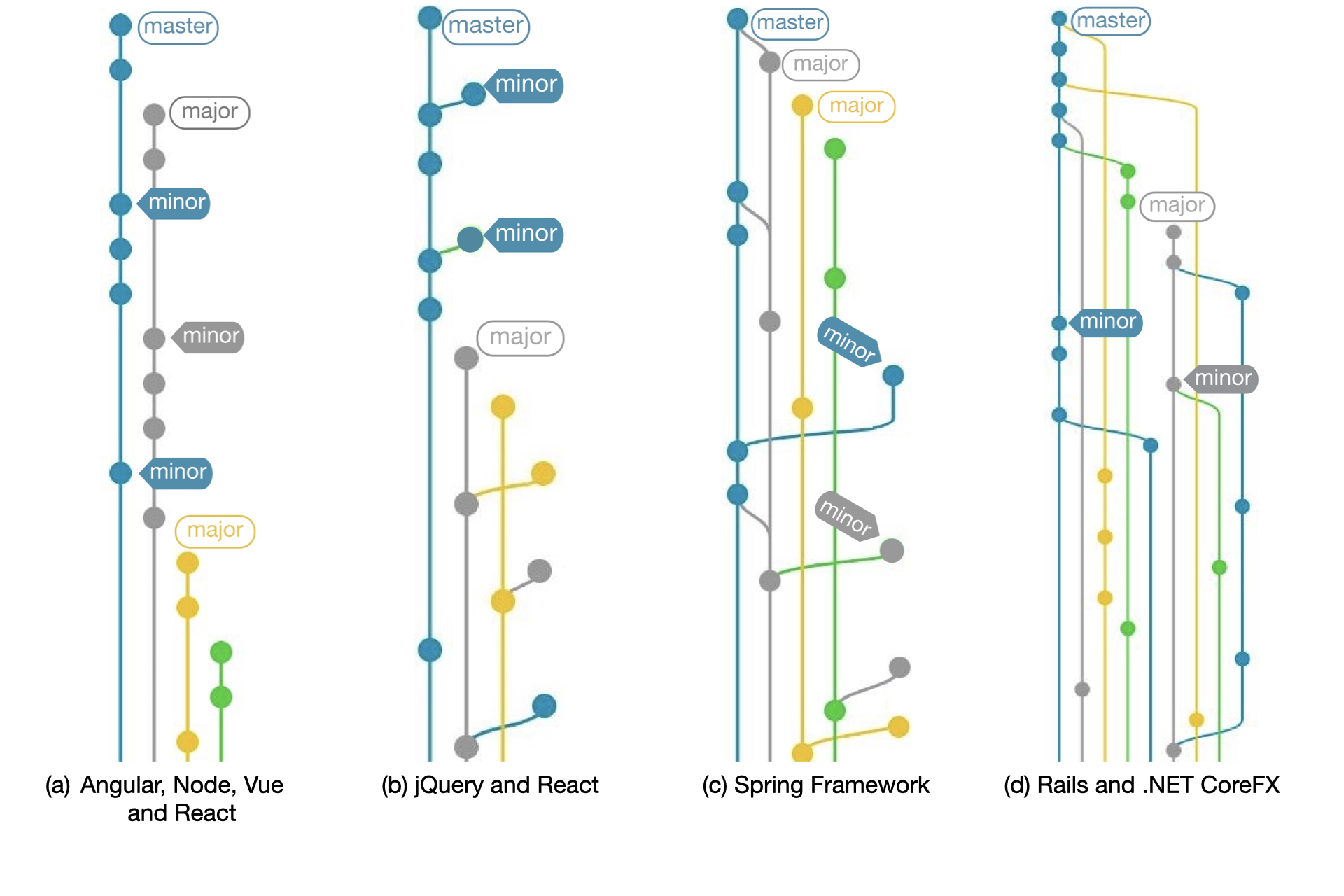}
\caption{four common patterns of commit graphs}
\end{figure*}

\end{document}